\RequirePackage{fix-cm}
\documentclass{svjour3}                     % onecolumn (standard format)
\smartqed  % flush right qed marks, e.g. at end of proof
\usepackage[acronym]{glossaries}
\usepackage[naturalnames]{hyperref}
\usepackage{graphicx}
\usepackage{caption}
\usepackage{subcaption}
\usepackage{booktabs}
\usepackage{threeparttable}
\usepackage{siunitx}
\usepackage[round]{natbib} % for APA like citation

\usepackage{multirow}

\makeatletter
\let\cl@chapter\undefined
\makeatletter

\newacronym{cp}{CP}{critical power}
\newacronym{w'}{\ensuremath{\mathrm{W}^\prime}}{finite energy reserve for work above critical power}
\newacronym{w'bal}{\ensuremath{\mathrm{W}^\prime}~balance}{work-balance}
\newacronym{hyd_weig}{\ensuremath{\mathrm{hydraulic}_\mathrm{weig}}}{hydraulic model by \cite{weigend_new_2021}}
\newacronym{m-m}{\mbox{M-M}}{Margaria-Morton}
\newacronym{m-m-s}{\mbox{M-M-S}}{Margaria-Morton-Sunstr\"om}
\newacronym{LAT}{LAT}{lactate threshold (moderate-heavy boundary)}

\newacronym{hyd_2t}{\ensuremath{\mathrm{hydraulic}_\mathrm{2t}}}{two-tank hydraulic}
\newacronym{SEE}{SEE}{standard error of estimation}
\newacronym{tte}{TTE}{time-to-exhaustion}
\newacronym{vo2}{\ensuremath{\dot{V}_{\mathrm{O}_2}}}{oxygen uptake}
\newacronym{vo2max}{\ensuremath{\dot{V}_{\mathrm{O}_2\mathrm{max}}}}{maximal oxygen uptake} % is an important determinant of their endurance capacity during prolonged exercise
\newacronym{vo2peak}{\ensuremath{\dot{V}_{\mathrm{O}_2\mathrm{peak}}}}{peak oxygen uptake}
\newacronym{p_peak}{$\mathrm{P}_\mathrm{peak}$}{peak power output}

\newacronym{AnA}{\ensuremath{A_nA}}{anaerobic alactic energy source}
\newacronym{AnL}{\ensuremath{A_nL}}{anaerobic lactic energy source}
\newacronym{O}{\ensuremath{O}}{oxidative or aerobic energy source}
\newacronym{Ae}{\ensuremath{Ae}}{aerobic energy source}
\newacronym{An}{\ensuremath{An}}{anaerobic energy sources}

\newacronym{AnS}{\ensuremath{AnS}}{anaerobic slow energy source}
\newacronym{AnF}{\ensuremath{AnF}}{anaerobic fast energy source}

\newacronym{m_Ae}{\ensuremath{m^{Ae}}}{maximal flow from \ensuremath{Ae}}
\newacronym{m_AnS}{\ensuremath{m^{AnS}}}{maximal flow from \ensuremath{AnS}}
\newacronym{m_AnF}{\ensuremath{m^{AnF}}}{maximal flow from \ensuremath{AnF}}

\usepackage{chngpage}
\usepackage{cleveref}

\journalname{Currently under Review}
\date{}

\begin{document}

\title{Benefits and limitations of a new hydraulic performance model}

% if too long for running head
%\titlerunning{Hydraulic model predictions for constant high intensity exercise}  

\author{
Fabian C. Weigend\hspace{1mm}\href{https://orcid.org/0000-0001-8868-9735}{\includegraphics[scale=0.06]{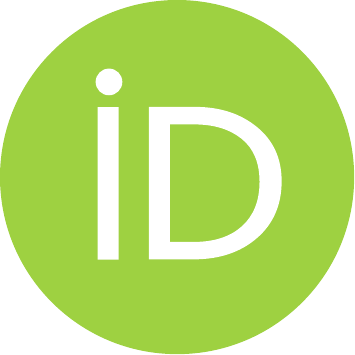}}
\and
Edward Gray\hspace{1mm}\href{https://orcid.org/0000-0003-2274-2147}{\includegraphics[scale=0.06]{fig_orcid.pdf}}
\and
Oliver Obst\hspace{1mm}\href{https://orcid.org/0000-0002-8284-2062}{\includegraphics[scale=0.06]{fig_orcid.pdf}}  
\and
Jason Siegler\hspace{1mm}\href{https://orcid.org/0000-0003-1346-4982}{\includegraphics[scale=0.06]{fig_orcid.pdf}}
}


\institute{
        Fabian C. Weigend \at
            School of Computer, Data and Mathematical Sciences and School of Health Sciences \\
            Western Sydney University, Sydney, Australia\\
            \email{Fabian.Weigend@westernsydney.edu.au}
        \and
        Edward Gray \at
            School of Health Sciences \\
            Western Sydney University, Sydney, Australia
        \and
        Oliver Obst \at
            School of Computer, Data and Mathematical Sciences\\
            Western Sydney University, Sydney, Australia
        \and
        Jason Siegler \at
            College of Health Solutions\\
            Arizona State University, Phoenix, USA
}


\maketitle

\begin{abstract}
\newline
\newline
\noindent\textbf{Purpose} Performance models are important tools for coaches and athletes to optimise competition outcomes or training schedules. A recently published hydraulic performance model has been reported to outperform established work-balance models in predicting recovery during intermittent exercise. The new hydraulic model was optimised to predict exercise recovery dynamics. In this work, we hypothesised that the benefits of the model come at the cost of inaccurate predictions of metabolic responses to exercise such as \gls{vo2}. 

\noindent\textbf{Methods} Hydraulic model predictions were compared to breath-by-breath \gls{vo2} data from 25 constant high-intensity exercise tests of 5 participants (age $32\pm7.8$ years, \mbox{weight $73.6 \pm 5.81$ kg}, \mbox{$\ensuremath{\dot{V}_{\mathrm{O}_2\mathrm{max}}} \; 3.59 \pm 0.62$ L/min}). 
Each test was performed to volitional exhaustion on a cycle ergometer with a duration between 2 and 12 min. The comparison focuses on the onset of \gls{vo2} kinetics. 

\noindent\textbf{Results} On average, the hydraulic model predicted peak $\dot{V}_{\mathrm{O}_2}$ during exercise $216\pm113$~s earlier than observed in the data. The new hydraulic model also did not predict the so-called \gls{vo2} slow component and made the unrealistic assumption that there is no \gls{vo2} at the onset of exercise.

\noindent\textbf{Conclusion} While the new hydraulic model may be a powerful tool for predicting energy recovery, it should not be used to predict metabolic responses during high-intensity exercise. The present study contributes towards a more holistic picture of the benefits and limitations of the new hydraulic model. Data and code are published as open source.

\end{abstract}

\vspace{1cm}
\noindent\textbf{Abbreviations} 

\begin{center}
\begin{tabular}{ l l }

CP & critical power \\ 
\ensuremath{\mathrm{W}^\prime} & finite energy reserve for work above critical power \\  \ensuremath{\mathrm{W}^\prime}~balance & work-balance model \\
TTE & time to exhaustion \\
SEE & standard error of estimation \\
LAT & lactate threshold (moderate-heavy boundary) \\
\ensuremath{\dot{V}_{\mathrm{O}_2}} & oxygen uptake \\
\ensuremath{\dot{V}_{\mathrm{O}_2\mathrm{max}}} & maximal oxygen uptake \\
\ensuremath{\dot{V}_{\mathrm{O}_2\mathrm{peak}}} & peak oxygen uptake during a single test \\
\ensuremath{\mathrm{P}_\mathrm{peak}} & peak power output \\

\mbox{M-M model} & Margaria-Morton model\\
\mbox{M-M-S model} & Margaria-Morton-Sunstr\"om model\\ 
\ensuremath{O} & oxidative or aerobic energy source\\
\ensuremath{A_nA} & anaerobic alactic energy source\\
\ensuremath{A_nL} & anaerobic lactic energy source\\
\ensuremath{T} & tap at the bottom of $A_nA$\\
\ensuremath{R1} & flow from \ensuremath{O}\\
\ensuremath{R2} & flow from \ensuremath{A_nL}\\
\ensuremath{R3} & flow from \ensuremath{A_nA}\\
$B$ & tube to account for early lactic acid occurrence\\
$\theta, \gamma, \phi$ & tank distances to bottom and top \\

$\mathrm{hydraulic}_{\mathrm{2t}}$ & hydraulic two-tank model \\
\ensuremath{Ae} & aerobic energy source\\
\ensuremath{An} & anaerobic energy sources\\
\ensuremath{p^{Ae}} & flow from $Ae$\\
\ensuremath{p} & tap to simulate power output \\

\ensuremath{\mathrm{hydraulic}_\mathrm{weig}} & hydraulic model by \cite{weigend_new_2021} \\
\ensuremath{AnF} & anaerobic fast energy source\\
\ensuremath{AnS} & anaerobic slow energy source\\
\ensuremath{m^{Ae}} & maximal flow from \ensuremath{Ae} \\
\ensuremath{m^{AnS}} & maximal flow from \ensuremath{AnS} \\
\ensuremath{m^{AnF}} & maximal flow from \ensuremath{AnF} \\

\ensuremath{U} & unlimited energy source\\
\ensuremath{LF} & limited fast energy source\\
\ensuremath{LS} & limited slow energy source\\
\ensuremath{M_U} & maximal flow from \ensuremath{U} \\
\ensuremath{M_{LF}} & maximal flow from \ensuremath{LF} \\
\ensuremath{M_{LS}} & maximal flow from \ensuremath{LS} \\
\end{tabular}
\end{center}

\section{Introduction}
\label{intro}

The quantification of performance (capacity) are a core requirement for optimising training and competition outcomes of an athlete. Performance models are tools for coaches and athletes to obtain such objective quantifications. One of the seminal models in this pursuit is the critical power model \citep{monod_work_1965,jones_maximal_2019}. The critical power model uses the parameters \gls{cp} and a \gls{w'}. Physiologically, \gls{cp} is defined as the threshold between heavy- and severe-intensity exercise \citep{jones_maximal_2019}. The capacity \gls{w'} limits the time an athlete can exercise at a severe intensity above \gls{cp}. \cite{hill_critical_1993} summarised the assumptions of the critical power model as the following:
\begin{itemize}
    \item[1.] An individual's power output is a function of two energy sources: aerobic (using oxidative metabolism) and anaerobic (non-oxidative metabolism).
    \item[2.] Aerobic energy is unlimited in capacity but its conversion rate into power output is limited (\gls{cp}).
    \item[3.] Anaerobic energy is limited in capacity but its conversion rate is unlimited.
    \item[4.] Exhaustion occurs when all of the anaerobic energy capacity is depleted.
\end{itemize}

\cite{whipp_constant_1982} then denoted the anaerobic energy capacity as \gls{w'} and both terms were used interchangeably. However, more recent publications suggest that they are not the same \citep{dekerle_validity_2006}. \cite{noordhof_determining_2013} summarised that \gls{w'} should not be considered as an entirely anaerobic entity and \cite{poole_critical_2016} suggested the conceptualisation of \gls{w'} as a buffer ``to resist exercise intolerance above \gls{cp}, where the source of the buffer will vary depending on the conditions". 

As already shown by varying definitions for anaerobic energy capacity and \gls{w'}, all of the above written four assumptions by \cite{hill_critical_1993} are incorrect from a strict physiological point of view \citep{morton_critical_2006,clarke_rationale_2013,poole_critical_2016}. Nevertheless, \citet{poole_critical_2016,jones_maximal_2019} described that the elegant abstraction of the critical power model proved to be useful for predicting energy expenditure of an athlete. 

Since then, performance models have evolved and a so-called \gls{w'bal} model has gained increasing popularity due to its intuitive combination of energy expenditure and recovery predictions. \cite{skiba_w_2021} summarised that, even though predictions of \gls{w'bal} models were of mixed quality, they convince because of their simplicity and their promise of tracking the energy capacities of an athlete in real-time. However, recent findings suggest that current \gls{w'bal} models overly simplify energy recovery dynamics \citep{bartram_accuracy_2018,caen_reconstitution_2019, caen_w_2021}. This is also reflected in reviews by \cite{jones_critical_2017,sreedhara_survey_2019,skiba_w_2021} who highlighted future work on \gls{w'bal} models to be of great promise for advances in performance modelling.

%\citet{hill_critical_1993, morton_critical_2006} summarised that the critical power and \gls{w'bal} models equate power output of an athlete as a function of two energy sources: a sustainable energy source for power output $<\gls{cp}$ and a limited energy source with \gls{w'} as its capacity for power output $>\gls{cp}$. \citet{morton_critical_2006} presented a hydraulic analogy for the critical power model.  performance models equate exercise performance within the paradigm of  metabolism. They are called hydraulic models because they represent human bioenergetic responses to exercise as liquid flow within a system of pipes and tanks. 

\subsection{A hydraulic analogy}

\begin{figure}
    \centering
    \includegraphics[width=0.4\textwidth]{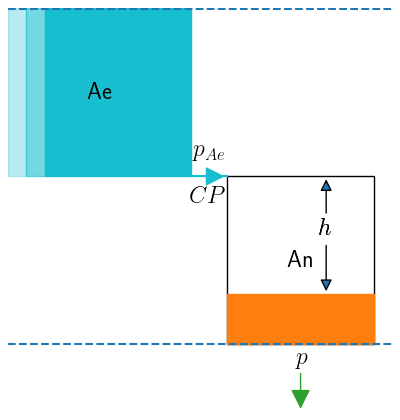}
    \caption{A hydraulic two-tank model as presented by \citet{morton_critical_2006}. The left tank represents the \gls{Ae} and is of infinite capacity, which is indicated by the fading colour to the left. The right tank represents \gls{An}. A pipe ($p_{\mathrm{Ae}}$) connects the bottom of \gls{Ae} to \gls{An} and has the maximal flow capacity \gls{cp}. A tap $p$ is attached to the bottom of \gls{An} and $h$ indicates the level of depletion of \gls{An}.} 
    \label{fig:two-tank}
\end{figure}

Next to critical power and \gls{w'bal} models, hydraulic models represent human bioenergetic responses to exercise as liquid flow within a system of pipes and tanks. \citet{morton_critical_2006} discussed a hydraulic model with two tanks as an analogy to the assumptions of the critical power model. Henceforth, it will be referred to as the \gls{hyd_2t} model. A schematic of the \gls{hyd_2t} model is depicted in \Cref{fig:two-tank}. 

The first of the above listed assumptions of the critical power model is that power output is a function of two energy sources: aerobic and anaerobic. Each energy source is represented by one tank in \Cref{fig:two-tank}. The second assumption says that aerobic energy is unlimited. Therefore, the tank \gls{Ae} is infinite in capacity, indicated by the fading colour to the left. The third assumption is that anaerobic energy is limited in its capacity. 

A pipe ($p_{Ae}$) connects the bottom of the aerobic tank to the anaerobic tank. The flow from this pipe represents sustainable energy contribution because the aerobic tank has infinite capacity. The second assumption states that the conversion of aerobic energy into power output is limited by \gls{cp}. Therefore, the pipe has the maximal flow capacity \gls{cp}. At the bottom of the \gls{An} tank is a tap $p$. Flow from this tap represents power output.

As discussed above, it is controversial whether \gls{w'} equates the anaerobic work capacity and, e.g. \citet{sreedhara_survey_2019} avoided this connotation and named the anaerobic tank "Limited capacity" when they described the \gls{hyd_2t} model. The present example specifically discusses the \gls{hyd_2t} model as an analogy to the above listed assumptions of the critical power model by \citet{hill_critical_1993}. Therefore, in this instance, it is justified to interpret the limited tank as anaerobic energy sources with capacity \gls{w'}. The fill-state of \gls{An} can drop and rise by $h$, and the remaining liquid represents available \gls{w'} balance.

At the beginning of exercise, the anaerobic tank is filled. If power output is below \gls{cp}, flow from the tap $p$ can be matched by flow from the aerobic tank, and the fill-level of the anaerobic tank does not drop. When power output rises above \gls{cp}, maximal flow from the aerobic tank is reached and the liquid level of the anaerobic tank drops by the difference between flow from $p$ and \gls{cp} at every time step. Exhaustion is reached when the anaerobic tank is depleted. Liquid flow within the example \gls{hyd_2t} model resembles the relation of power output, \gls{cp}, and \gls{w'} as assumed by the critical power model. 

\subsection{The M-M model}

\begin{figure}
    \centering
    \includegraphics[width=\textwidth]{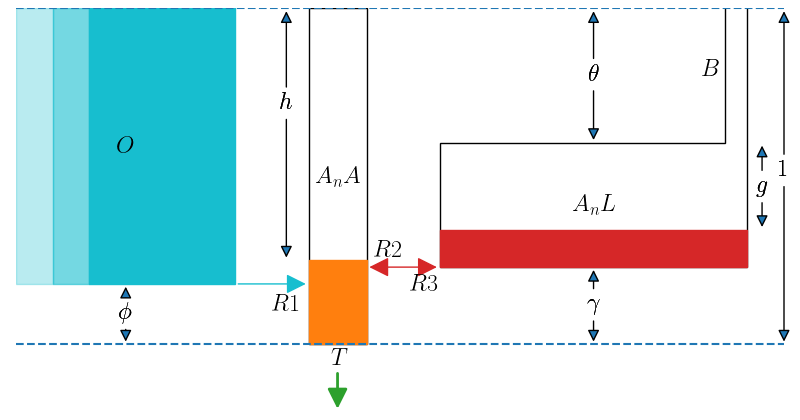}
    \caption{The \gls{m-m} model as published by \cite{morton_critical_2006}. Hydraulic models approximate human bioenergetic responses to exercise as liquid flow within a system of pipes and tanks. The left infinitely big tank $O$ represents the oxidative or aerobic energy source. The limited tank in the middle ($A_nA$) represents the alactic anaerobic phosphagens. The right limited tank ($A_nL$) represents an anaerobic lactic energy source. A tap ($T$) is attached to $A_nA$, which can be opened and closed according to energy demand. The tube $B$ accounts for early lactic acid occurrence. The pipes $R_1$, $R_2$, $R_3$ enable flow between the tanks. Tank sizes are defined by $\theta, \gamma, \phi$.} 
    \label{fig:morton_m_m_2006}
\end{figure}

Next to \gls{hyd_2t}, more complex hydraulic models make use of liquid pressure dynamics and make predictions for metabolic responses during exercise with three or four tanks \citep{morton_critical_2006,sundstrom_bioenergetic_2016}. The first hydraulic model was published by \cite{margaria_biomechanics_1976} and later further elaborated by \cite{morton_critical_2006}. Morton mathematically defined its dynamics and published it as the \gls{m-m} model. An example of the \gls{m-m} model with the notation of \cite{morton_critical_2006} is depicted in \Cref{fig:morton_m_m_2006}. Like \cite{margaria_biomechanics_1976}, \cite{morton_critical_2006} labelled the left infinitely big tank as \gls{O}. Similarly to \gls{Ae} of \gls{hyd_2t}, the tank \gls{O} has infinite capacity. The limited tank in the middle represents the \gls{AnA} and the third limited tank on the right represents the \gls{AnL} \citep{morton_critical_2006}. $\theta, \gamma$ and $\phi$ define tanks sizes and affect liquid flow dynamics. $\theta$ is the distance between the top of \gls{AnA} and the top of \gls{AnL}. $\phi$ and $\gamma$ are the distances between the bottoms of \gls{O} to \gls{AnL} and to \gls{AnA} respectively. The pipes $R_1$, $R_2$, $R_3$ enable flow between the tanks and have maximal flow capacities.

The depicted situation in \Cref{fig:morton_m_m_2006} is that $T$ has been opened and the fill-level of the middle tank $A_nA$ dropped by $h$. The more liquid that flows out of the middle tank, the lower the liquid pressures against $R_1$, and the more flows from $O$ through $R_1$ into $A_nA$. In this way $R_1$ contributes to the flow out of $T$. In the situation depicted in \Cref{fig:morton_m_m_2006}, $T$ was opened so wide that $h > \theta$ and flow from $A_nL$ began to contribute as well. Its fill-level had dropped by $g$.

Because the pipe $R_1$ allows flow from the aerobic source $O$ into the middle tank, \cite{morton_critical_2006} referred to flow through $R_1$ as \gls{vo2}. He also defined that the maximal flow through pipe $R_1$ represents the \gls{vo2max}. \cite{morton_three_1986} fitted differential equations of his model to collected \gls{vo2} dynamics and the model could explain the observations well. However, \cite{morton_critical_2006} concluded it had yet to be seen what predictions of the \gls{m-m} model conform to reality because the model parameters are extremely difficult or impossible to obtain from individual athletes, e.g., $\theta, \gamma, \phi$ have no direct physiological analogy and the capacities of \gls{AnL} and \gls{AnA} can only be approximated. 

\cite{sundstrom_comparing_2014} investigated predictions of the \gls{m-m} model in theoretical elaborations. They compared predicted optimal pacing strategies of the \gls{m-m} model to those of a critical power model for intermittent exercise on an artificial course and reported that the \gls{m-m} model made more realistic predictions. \cite{sundstrom_bioenergetic_2016} also developed a sophistication of the \gls{m-m} model and published it as the \gls{m-m-s} model. However, their work was theoretical, and, like \citet{morton_critical_2006}, they also concluded that both their findings and model had yet to be validated on real athlete data. 

\citet{lidar_validity_2021} developed an approach to validate the \gls{m-m} model on real athlete data by fitting parameters to available measurements with an optimisation approach. They fitted two versions of the \gls{m-m} model and two \gls{hyd_2t} models to measured aerobic metabolic rate and accumulated anaerobic energy expenditure during treadmill roller-skiing time trials. \citet{lidar_validity_2021} reported that the \gls{hyd_2t} model provided the highest validity and reliability for data it was fitted to and for predictions on unknown data. Further, \citet{lidar_validity_2021} observed that optimal parameters for the fitted \gls{m-m} model were likely outside the physiologically reasonable ranges and they concluded that the \gls{m-m} model cannot fully capture bioenergetic responses of the human body to exercise.

\subsection{Hydraulic models compared to \gls{w'bal} models}

In \citet{weigend_new_2021}, we proposed an alternative hydraulic model that, in contrast to \citet{morton_critical_2006, sundstrom_bioenergetic_2016, lidar_validity_2021}, does not ascribe hydraulic model parameters to alactic or lactic energy sources. The model will henceforth be referred to as \gls{hyd_weig}.

\begin{figure}
    \centering
    \includegraphics[width=\textwidth]{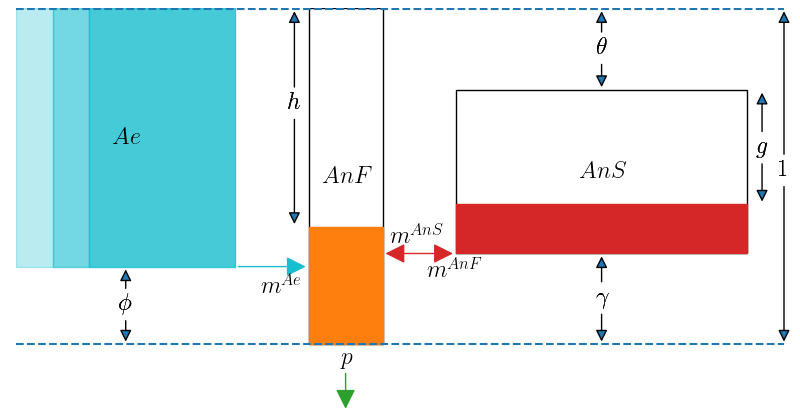}
    \caption{Representative example of the hydraulic model by \gls{hyd_weig} as published in \cite{weigend_new_2021}. In comparison to the \gls{m-m} model in \Cref{fig:morton_m_m_2006}, tube $B$ was removed and the tanks have been renamed as aerobic energy source ($Ae$), anaerobic fast energy source ($AnF$) and anaerobic slow energy source ($AnS$). The maximal flow capacities from tanks are labelled as $m^{Ae}$, $m^{AnS}$, $m^{AnF}$. Tank sizes are defined by $\theta, \gamma, \phi$.} 
    \label{fig:weigend_old}
\end{figure}

%\Gls{hyd_weig} represents a different take on hydraulic models because it is not intended to make predictions for metabolic responses to exercise, e.g, for alactic or lactic energy systems. Rather, it 
\Gls{hyd_weig} is intended to predict energy expenditure and recovery during high-intensity intermittent exercise in a more general sense. This is reflected in the schematic of \gls{hyd_weig} depicted in \Cref{fig:weigend_old}. Tube $B$ was removed and the middle tank was re-imagined as general anaerobic fast energy sources ($AnF$) and the right tank as anaerobic slow energy sources ($AnS$). These changes removed physiological constraints that \citet{morton_critical_2006, sundstrom_bioenergetic_2016, lidar_validity_2021} had to apply to their models and allowed us to adjust and interpret model parameters more freely.

The removal of physiological constraints allowed \gls{hyd_weig} to serve as an alternative to \gls{w'bal} models for real-time \gls{tte} predictions during high-intensity intermittent exercise. \citet{weigend_new_2021} proposed a pathway to fit \gls{hyd_weig} to \gls{cp} and \gls{w'} of an athlete. These are the same inputs required by \gls{w'bal} models. When given these inputs, an evolutionary algorithm fits parameters of \gls{hyd_weig} such that it predicts \gls{tte} during constant high-intensity exercise according to the critical power model, and such that it predicts recovery of \gls{w'} according to measurements derived from a study by \cite{caen_reconstitution_2019}. Specifically, that means the parameters of a fitted \gls{hyd_weig} model do not directly correspond to \gls{cp} or \gls{w'} but the model was fitted to make predictions similar to the critical power model and to published observations for energy recovery. 

Therefore, \gls{hyd_weig} represents a new approach for how hydraulic models can be used. % Original models like the \gls{m-m} model and \gls{m-m-s} focus on metabolic responses during exercise, while \gls{hyd_weig} abstains from ascribing parameters to alactic or lactic energy sources or to \gls{cp} and \gls{w'}. \Gls{hyd_weig} predicts energy expenditure and recovery in a more general sense and is intended to be used for more accurate recovery predictions during intermittent exercise. 
In \cite{weigend_hydraulic_2021}, we retrospectively compared energy recovery predictions of \gls{w'bal} and \gls{hyd_weig} models with published data from five studies. The prediction capabilities of \gls{hyd_weig} outperformed \gls{w'bal} models on all metrics and rendered it as a strong direction for future research in performance modelling.  

However, a remaining similarity between \gls{hyd_weig} and the \gls{m-m} model is that the left tank is still labelled as the aerobic energy source. This indicates that the flow from \gls{Ae} still represents \gls{vo2}. Because \citet{lidar_validity_2021} recently concluded that the \gls{m-m} model cannot completely capture the human bioenergetic system and because the parameters of \gls{hyd_weig} are fitted without physiological constraints, we hypothesised that flow from \gls{Ae} of a fitted \gls{hyd_weig} does not predict realistic \gls{vo2} dynamics. This had not yet been investigated, and therefore we designed this work to confirm the limitations of \gls{hyd_weig}. Our goal is to support users in interpreting \gls{hyd_weig} correctly and make users aware of its benefits, but also of its limitations. 

\section{Methodology}

To verify \gls{hyd_weig} limitations, we compared \gls{vo2} predictions of \gls{hyd_weig} to collected \gls{vo2} data of exercising participants. \gls{cp} and \gls{w'} are required to apply \gls{hyd_weig}. These parameters are estimated from \gls{tte} performance tests during which participants exercise at constant high intensities until volitional exhaustion. Therefore, we scrutinised \gls{hyd_weig} \gls{vo2} predictions on collected data from these \gls{tte} tests. 

\subsection{Data collection}

Data collection was approved by the Human Research Ethics Committee at Western Sydney University (HREC Approval Number: H13975). Five recreationally active participants (4 males and 1 female, age $32\pm7.8$~years, weight $73.6 \pm 5.81$~kg, $\ensuremath{\dot{V}_{\mathrm{O}_2\mathrm{max}}} = 3.59 \pm 0.62$~L/min) gave informed and written consent to participate and to have their anonymised data published. All participants were familiar with maximal exercise efforts. Exercise tests were conducted on an SRM - High Performance Ergometer (Jülich, Germany) in hyperbolic operation mode, which adjusts power dynamically to cadence changes to maintain a constant power output. Breath-by-breath \gls{vo2} data was collected using the Quark CPET system by COSMED (Rome, Italy). The equipment was calibrated prior to each trial. Each of the 5 participants completed 6 exhaustive exercise trials. To ensure that participants were fully rested, they were asked to avoid strenuous exercise 24~h prior the tests and tests were scheduled more than 24~h apart, roughly at the same time of the day. 

\subsubsection{Ramp test}

All participants had to perform an initial ramp test to obtain the appropriate power settings for subsequent \gls{tte} tests. After a 3-min warm-up at 50~W, the power increased by 30~W per min for males and by 20~W per min for the female. Once power rose above 110~W, participants were instructed to maintain a self-chosen cadence between 80-100~RPM. The point of volitional exhaustion was defined as the first time point when the cadence dropped by more than 10\% below the intended cadence for more than 3~s. The highest 10-s moving average power output achieved was defined as the \gls{p_peak}. 

\subsubsection{TTE tests}

After the ramp test, each participant completed 5~constant power \gls{tte} trials at distinct powers in random order. The powers were set to 100\%, 92.5\%, 85\%, 80\% and 77.5\% or 75\% of $\mathrm{P}_\mathrm{peak}$ to obtain a range of \gls{tte}s between 2~min and 12~min. Participants were blinded to their exercise power. Again, each test started with a 3-min warm-up at 50~W before power was set to the randomly chosen percentage of $\mathrm{P}_\mathrm{peak}$. During exercise, participants were asked to cycle at their self-chosen cadence from the ramp test and the point of volitional fatigue was defined as the first time point when cadence dropped by more than 10\% below the intended cadence for more than 3~s. Throughout all tests, breath-by-breath \gls{vo2} data was collected. For each test, the highest achieved 30-s moving average of measured \gls{vo2} was considered as the \gls{vo2peak} of that test. \gls{tte}s as well as power-meter data of the SRM ergometer were also recorded for later analysis.

\subsection{Data analysis}

\citet{weigend_new_2021} designed \gls{hyd_weig} to serve as an alternative to \gls{w'bal} models and---like \gls{w'bal} models---\gls{hyd_weig} requires \gls{cp} and \gls{w'} to make predictions. These parameters were obtained by fitting the critical power model to conducted \gls{tte} tests of a participant.

\subsubsection{Fitting the models}

To obtain \gls{cp} and \gls{w'} that best fit a participant, three forms of the critical power model were fitted to conducted \gls{tte} tests. These forms were
\begin{equation}
    \gls{tte} = \frac{\gls{w'}}{P - \gls{cp}},
\end{equation}
which will be henceforth referred to as the nonlinear power-time model,
\begin{equation}
    P \cdot \gls{tte} = \gls{cp} \cdot \gls{tte} + \gls{w'},
\end{equation}
which will be henceforth referred to as the linear work-time model, and
\begin{equation}
    P = \gls{w'} \cdot \frac{1}{\gls{tte}} + \gls{cp},
\end{equation}
which will be henceforth referred to as the linear power-1/time model \citep{hill_critical_1993}. 

The goodness of fit of each model was determined from the \gls{SEE} associated with fitted \gls{cp} and \gls{w'}. The goodness of fit of a model was considered sufficient if \gls{SEE} associated with \gls{cp} was $<5\%$ of \gls{cp} and the \gls{SEE} associated with \gls{w'} was $<10\%$ of \gls{w'} \citep{jones_maximal_2019,caen_w_2021}. The best individual fit for a participant was selected by which model resulted in the smallest sum of the \gls{SEE} associated with \gls{cp} as \% of \gls{cp} plus the \gls{SEE} associated with \gls{w'} as \% of \gls{w'} \citep{black_self-pacing_2015,jones_maximal_2019,caen_w_2021}.

Then, \gls{hyd_weig} was fitted to the derived \gls{cp} and \gls{w'} with the procedure published in \cite{weigend_new_2021}. Considering the notation in \Cref{fig:weigend_old}, fitting \gls{hyd_weig} meant finding a set of parameters for $[ AnF, AnS, \allowbreak m^{Ae}, m^{AnS}, \allowbreak m^{AnF}, \theta, \gamma, \phi ]$ that made the hydraulic model resemble expected exercise responses according to the critical power model for energy expenditure, and according to published recovery ratios for energy recovery \citep{caen_reconstitution_2019, weigend_new_2021}. To find these parameters for our participants, we used the automatised evolutionary computation procedure of our \texttt{threecomphyd}\footnote{\url{https://github.com/faweigend/three_comp_hyd}} Python package, which was published in \cite{weigend_new_2021}. 

\subsubsection{\gls{vo2} predictions}

\begin{figure}
    \makebox[\textwidth][c]{
    \begin{minipage}{0.45\textwidth}
        \centering
        \includegraphics[width=\textwidth]{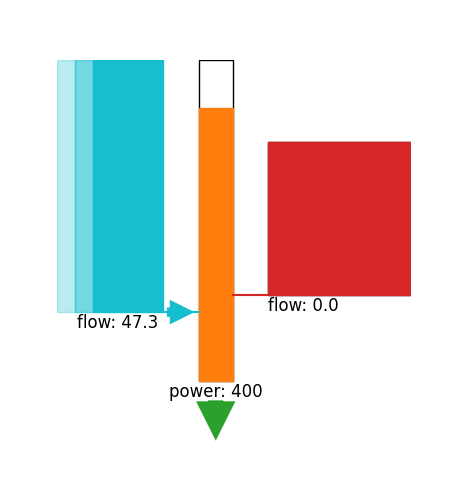}
        \subcaption{10~s}
    \end{minipage}\hfill
    \begin{minipage}{0.45\textwidth}
        \centering
        \includegraphics[width=\textwidth]{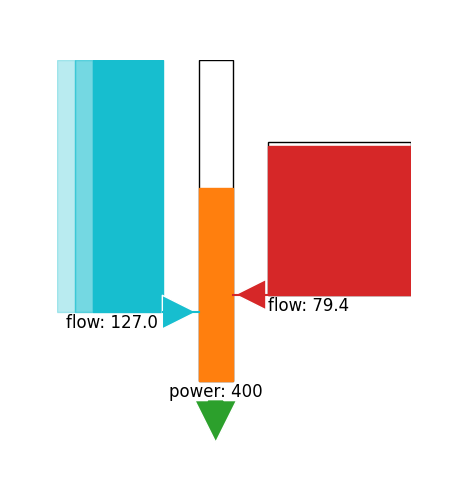}
        \subcaption{20~s}
    \end{minipage}\hfill
        \begin{minipage}{0.45\textwidth}
        \centering
        \includegraphics[width=\textwidth]{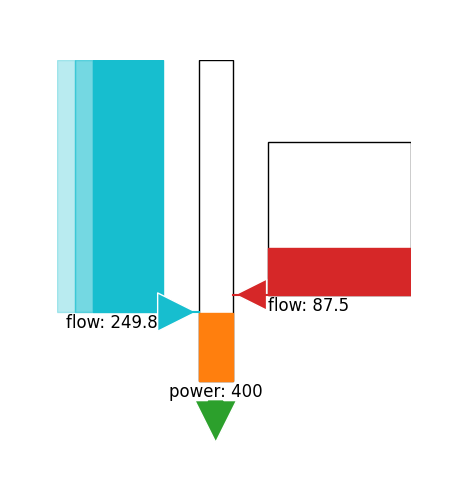}
        \subcaption{80~s}
    \end{minipage}}
    \caption{Snapshots of a \gls{hyd_weig} model simulating exercise at a constant intensity of 400 W. Depicted are fill-levels of tanks and the flows between them after 10, 20, and 80~s. 
    %The parameters of \gls{hyd_weig} in this example are $ AnF=11532\;\mathrm{J}, AnS=23240\;\mathrm{J}, m^{Ae}=249\;\mathrm{W}, m^{AnS}=286\;\mathrm{W}, m^{AnF}=8\;\mathrm{W}, \theta=0.25, \gamma=0.27, \phi=0.21 $.
    } \label{fig:vo2_sim_example}
\end{figure}

To assess the quality of predicted \gls{vo2} kinetics of fitted \gls{hyd_weig} models, the predictions were compared to the collected breath-by-breath \gls{vo2} data of the \gls{tte} tests. To elaborate how \gls{vo2} predictions were computed, we use the example in \Cref{fig:vo2_sim_example}, which displays snapshots of a \gls{hyd_weig} model that simulates an exercise with constant intensity. 

At the beginning of exercise at second 0, all tanks were filled. When exercise started, the tap at the bottom of the middle tank was opened according to the energy demand. Liquid flowed out, and the fill-level of the middle tank dropped accordingly. As a consequence, liquid pressure against the pipe exit of the left tank dropped and thus the liquid flow from the left tank increased. 

In the example in \Cref{fig:vo2_sim_example}, the tap was opened according to 400~W and after 10~s, the fill-level of the middle tank dropped halfway to the top of the right tank. Flow from the left tank increased to an equivalent of 47.3~W. As observable in the snapshot at 20~s, when the fill-level of the middle tank dropped below the top of the right tank, liquid from the right tank also flowed into the middle tank. In the snapshot at 80~s, the fill-level of the middle tank dropped below the exit of the pipe of the left tank, and thus the flow from the left tank was at maximum capacity $m^{Ae}$. 

According to the definitions of \cite{morton_critical_2006}, the left tank represents the aerobic contribution, and therefore the flow from the left tank represents oxygen uptake (\gls{vo2}). With this understanding, we opened and closed the tap according to collected SRM power-meter data and recorded flow from the left tank as predicted \gls{vo2} kinetics. We compared \gls{vo2} predictions with collected breath-by-breath data from all constant power \gls{tte} trials. 

Our objective measure to assess the quality of \gls{vo2} predictions was the difference between the time at which the simulated flow from the left tank ($Ae$) was predicted to reach its peak, and the time at which the observed breath-by-breath \gls{vo2} data reached \gls{vo2peak}. As an additional visual comparison, we plotted normalised predicted flow from $Ae$ together with normalised actual \gls{vo2} dynamics. Predicted \gls{vo2} dynamics (flow from $Ae$) were normalised with the maximal flow $m^{Ae}$. The 30~s averaged real \gls{vo2} uptake measurements were normalised with the observed \gls{vo2peak} during that test. 

\section{Results}
\label{sec:results}

\subsection{Ramp test results and model fittings}

The average $\mathrm{P}_\mathrm{peak}$ of the ramp tests of all participants was $327 \pm 52$ W. The shortest \gls{tte} was excluded from the estimation of \gls{cp} and \gls{w'} for one participant because it was too short (113~s). For all participants, the linear power-1/time model resulted in the best individual fit and resulted in an averaged \gls{cp} of $223\pm40$~W and \gls{w'} of $148912\pm2869$~J. Individual critical power model fitting results and associated \gls{SEE}s are summarised in \Cref{tab:cp_fitting_results} in the Appendix. 

Using the notation of \Cref{fig:weigend_old}, the fitted \gls{hyd_weig} models had an average $AnF$ of $14330 \pm 2463$~J, $AnS$ of $38575\pm6605$~J, $m^{Ae}$ of $222 \pm 40$~W, $m^{AnS}$  of $90 \pm 15$~W, $\theta$ of $0.7 \pm 0.05$, $\gamma$ of $0.02 \pm 0.01$, and $\phi$ of $0.26 \pm 0.04$. Individual results for these parameters are summarised in \Cref{tab:hyd_results} in the Appendix.

\subsection{\gls{vo2} predictions}

\begin{figure}
    \centering
    \includegraphics[width=\textwidth]{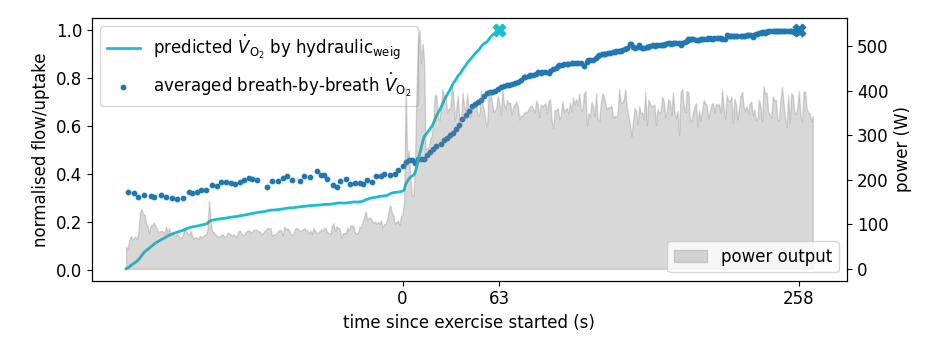}
    \caption{Measured power-meter output and \gls{vo2} of participant 2 during warm-up and a \gls{tte} test at 364~W. Measurements before second~0 are part of the 3-min warm-up at 50~W. The plotted line displays predicted \gls{vo2} uptake (flow from $Ae$) of a \gls{hyd_weig} model fitted to \gls{cp} and \gls{w'} of participant 2. The averaged collected breath-by-breath \gls{vo2} data is depicted as dots. The x-symbols mark the time points when \gls{hyd_weig} predicted \gls{vo2peak} (63~s) and when collected averaged breath-by-breath reached \gls{vo2peak} (258~s). The prediction error was 195~s.}
    \label{fig:tte}
\end{figure}

Of all 25 constant power tests, the tests of participant~4 at 288~W and participant~5 at 255~W were excluded from the \gls{vo2} analysis. In these two cases, unrealistic drops in \gls{vo2} indicated that the breathing mask was not fixed tight enough and leaked air when participants lowered their head too far. The unrealistic drops are clearly recognisable and depicted in \Cref{fig:vo2-03,fig:vo2-04} in the Appendix.

The defined objective measure to assess the quality of \gls{vo2} predictions was the time difference between predicted and obesrved \gls{vo2peak}. Furthermore, normalised flow from \gls{Ae} and measured breath-by-breath data were plotted for visual comparison. An example for such a plot is depicted in \Cref{fig:tte}. The power output measured by the SRM power-meter is plotted as the grey area in the background. It is overlayed with predicted and observed \gls{vo2} kinetics. The time at which \gls{hyd_weig} predicted \gls{vo2peak} was 63~s after the commencement of exercise. The time at which the actual \gls{vo2peak} was observed was 258~s after the commencement of exercise. Therefore, the prediction error was 195~s. In addition, it is observable that the predicted \gls{vo2} started at 0 and increased slowly during the warm-up.

\begin{table}[] 
\caption{The summary of \gls{hyd_weig} prediction errors for \gls{vo2peak}. The example from \Cref{fig:tte} (participant~2 with 364~W) is in row 8. The column ``observed'' informs about the seconds it took from the onset of exercise to reach \gls{vo2peak} (blue x-symbol at 258~s in \Cref{fig:tte}). The column ``predicted'' informs about the seconds \gls{hyd_weig} predicted it would take (azure x-symbol at 63~s in \Cref{fig:tte}). The prediction error in the last column is the difference between these two times.}
\label{tab:vo2_detail} 
\begin{adjustwidth}{}{}
\centering
\begin{tabular}{S[table-format=3] S[table-format=3] S[table-format=3] S[table-format=3] S[table-format=3]}
\toprule
 & & \multicolumn{2}{c}{time until \gls{vo2peak} (s)} & \multicolumn{1}{c}{}\\
\cmidrule(l){3-4}
\multicolumn{1}{c}{participant} & \multicolumn{1}{c}{power (W)} & \multicolumn{1}{c}{observed} & \multicolumn{1}{c}{predicted} & 
\multicolumn{1}{c}{prediction error (s)}\\
\midrule
1 & 243 & 481 & 72 & 409 \\
1 & 259 & 230 & 53 & 177 \\
1 & 275 & 196 & 45 & 151 \\
1 & 299 & 156 & 41 & 115 \\
1 & 324 & 120 & 34 & 86 \\
2 & 332 & 484 & 88 & 396 \\
2 & 343 & 297 & 78 & 219 \\
2 & 364 & 258 & 63 & 195 \\
2 & 396 & 169 & 54 & 115 \\
2 & 428 & 137 & 44 & 93 \\
3 & 252 & 479 & 113 & 366 \\
3 & 265 & 355 & 85 & 270 \\
3 & 280 & 230 & 72 & 158 \\
3 & 307 & 154 & 53 & 101 \\
3 & 330 & 109 & 42 & 67 \\
4 & 224 & 581 & 120 & 461 \\
4 & 230 & 445 & 94 & 351 \\
4 & 245 & 330 & 85 & 245 \\
4 & 265 & 266 & 67 & 199 \\
5 & 214 & 309 & 68 & 241 \\
5 & 221 & 292 & 55 & 237 \\
5 & 234 & 249 & 44 & 205 \\
5 & 276 & 152 & 32 & 120 \\
\midrule
 \multicolumn{2}{c}{avg$\pm$std}  & 
 \multicolumn{1}{c}{282$\pm$134} &
 \multicolumn{1}{c}{65$\pm$24} &   
 \multicolumn{1}{c}{216$\pm$113}\\
 \bottomrule \\
\end{tabular} 
\end{adjustwidth} 
\end{table}

The example \Cref{fig:tte} is representative for all tests. As summarised in \Cref{tab:vo2_detail}, on average, \gls{vo2peak} was observed after 282$\pm$134~s of exercise, while \gls{hyd_weig} predicted \gls{vo2peak} after 65$\pm$24~s of exercise in the respective test. In all the tests investigated, \gls{hyd_weig} predicted a much faster rise in \gls{vo2} and a too early \gls{vo2peak} with an average prediction error of $216 \pm 113$~s. \Cref{tab:vo2_detail} summarises our results.  The best \gls{hyd_weig} prediction was $67$~s too early. The worst prediction was $461$~s too early. The prediction error decreased as the exercise power increased.

\section{Discussion}

Previously, theoretical hydraulic performance models of \cite{morton_critical_2006} and \cite{sundstrom_bioenergetic_2016} promised predictions for metabolic responses during exercise, e.g, for lactic, alictic, and aerobic energy sources. But their models were not suitable for real-world applications because the required parameters to apply these models, e.g, precise lactic energy capacities in Joules, were impossible to obtain from individual athletes. \citet{lidar_validity_2021} fitted parameters for the \gls{m-m} model to available measurements and concluded that it likely cannot fully capture the human bioenergetic system. 

In \cite{weigend_new_2021}, we presented a more abstract hydraulic model that represented a different pathway for how hydraulic models can be used. \Gls{hyd_weig} was designed to serve as an alternative to \gls{w'bal} models for \gls{tte} predictions during intermittent exercise. To be comparable to \gls{w'bal} models \cite{weigend_new_2021} proposed a pathway to obtain \gls{hyd_weig} predictions from \gls{cp} and \gls{w'}. We observed in \cite{weigend_hydraulic_2021} that hydraulic energy recovery predictions outperformed energy recovery predictions by \gls{w'bal} models and our findings marked hydraulic models as a strong direction for future research in performance modelling. 

\Gls{hyd_weig} is a more general interpretation of the \gls{m-m} model. We hypothesised that, although the parameters of the \gls{hyd_weig} model have an aerobic and anaerobic connotation, it cannot make realistic predictions for \gls{vo2}. \Gls{hyd_weig} was designed to predict \gls{tte}s during intermittent exercise, but because its tanks and pipes resemble the \gls{m-m} model so closely, it is tempting to interpret \gls{hyd_weig} in a metabolic context. In the present study, we confirmed, with data collected from 23 performance tests of 5 participants, that \gls{hyd_weig} is not suitable for \gls{vo2} predictions. 

\subsection{Predictions of \gls{vo2} slow component}
\label{subsec:vo2_mm}

From the onset of high-intensity exercise, \gls{hyd_weig} consistently predicted \gls{vo2} to rise too fast. As summarised in \Cref{tab:vo2_detail}, \gls{vo2} was predicted to reach its peak after an average of 65$\pm$24~s while observed kinetics were slower and took 282$\pm$134. The average difference between \gls{vo2peak} predictions and observations was $216 \pm 113$~s. Considering that \gls{tte} tests lasted between 2-12 min, prediction errors of more than 3 min on average made clear that \gls{hyd_weig} could not predict realistic \gls{vo2} kinetics. 

Further, it is observable in \Cref{tab:vo2_detail} that the prediction error increases with decreasing power. Thus, the longer the exercise, the larger the error in predicted \gls{vo2}. These results are in contrast to remarks by \cite{morton_critical_2006} for \gls{vo2} prediction capabilities of his \gls{m-m} model and we elaborate the reasons for such poor predictions in the following. We begin by summarising the physiological constraints that \cite{morton_critical_2006} applied to his \gls{m-m} model.

\subsubsection{\gls{m-m} model and \gls{vo2} slow component}
\label{subsubsec:mm_slow_comp}

Looking at the \gls{m-m} model, tank positions are determined by the values $\theta, \gamma$ and $\phi$. Because each tank represents a concrete bioenergetic energy source, \cite{morton_three_1986,morton_modelling_1990,morton_critical_2006} developed several constraints on $\theta, \gamma$ and $\phi$ to find a realistic arrangement of tanks for his model:
\begin{itemize}
    \item Pipe $R_2$/$R_3$ has to be above $R_1$ ($\gamma > \phi$) because athletes can deplete their glycogen stores when exercising below \gls{vo2max}.
    \item The onset of flow through $R_2$ represents the \gls{LAT}, i.e., the commencement of increased lactic acid production. Therefore, the top of tank $A_nA$ has to have some distance to the top of the entire system ($\theta > 0$) and should be at approximately 40\% of the height of $O$.
    \item During constant severe intensity exercise, \gls{vo2} rises asymptotic to a maximal value. When exercise stops, oxygen consumption does not decline immediately. Therefore, $R_1$ cannot be at the top or bottom of the middle tank ($0 < \phi < 1$).
\end{itemize} 
From these constraints, \citet{morton_modelling_1990} argued that the only realistic configuration of the three component hydraulic model is the one depicted in \Cref{fig:morton_m_m_2006}.

\begin{figure}
    \makebox[\textwidth][c]{
    \begin{minipage}{0.45\textwidth}
        \centering
        \includegraphics[width=\textwidth]{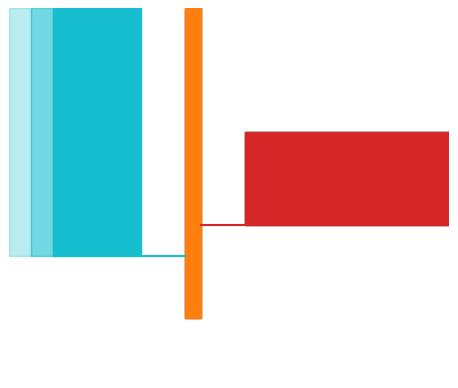} 
    \end{minipage}\hfill
    \begin{minipage}{0.45\textwidth}
        \centering
        \includegraphics[width=\textwidth]{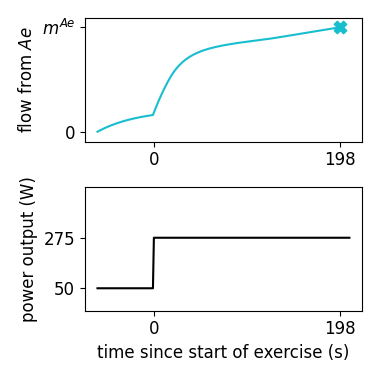}
    \end{minipage}}
    \caption{Left: A \gls{hyd_weig} model configured in a way that follows the physiological constraints on the \gls{m-m} model discussed in \Cref{subsubsec:mm_slow_comp}. Right: The top plot displays flow from the left tank ($Ae$) into the middle tank ($AnF$). The displayed model simulates constant intensity exercise at 275 W with a warm-up at 50 W until maximal flow through $Ae$ is reached (x-symbol). We interpreted this flow as \gls{vo2} uptake predictions. The \gls{vo2} slow component is observable in predicted \gls{vo2} dynamics.}
    \label{fig:m_m_sim}
\end{figure}

In his review, \cite{morton_critical_2006} particularly highlighted how the \gls{m-m} model predicts the \gls{vo2} slow component phenomenon. \cite{barstow_linear_1991} empirically showed that \gls{vo2} uptake quickly reaches a steady state at a constant exercise intensity below \gls{LAT}. However, at exercise above \gls{LAT} an initial rapid increase in \gls{vo2} uptake is followed by a slower continuous rise. This slower rise is called the \gls{vo2} slow component. 

As observable in \Cref{fig:m_m_sim}, the slow component is well captured by a hydraulic model that is configured according to the above stated constraints on the \gls{m-m} model. Depicted on the right in \Cref{fig:m_m_sim} are the predicted \gls{vo2} dynamics (flow from $Ae$) during constant high intensity exercise. With all tanks filled at the beginning, the dynamics play out as follows: During the warm-up, the tap was not opened wide and thus liquid level in the middle tank drops slowly and flow from $Ae$ increases slowly. Then exercise starts after one min, the fill-level in the middle tank drops quickly and therefore flow from $Ae$ increases quickly. The exercise intensity is high enough, so that the dropping fill-level of the middle tank reaches the top of the right tank, the right tank also starts to contribute. This additional flow makes the fill-level of the middle tank drop slower and therefore, flow from $Ae$ increases slower from this point. \cite{morton_critical_2006} stated that the way in which the \gls{m-m} model simulates the \gls{vo2} slow component can be compared to the mathematical formulation by \cite{barstow_linear_1991}. 

\subsubsection{\Gls{hyd_weig} and the \gls{vo2} slow component}

\begin{figure}
    \makebox[\textwidth][c]{
    \begin{minipage}{0.45\textwidth}
        \centering
        \includegraphics[width=\textwidth]{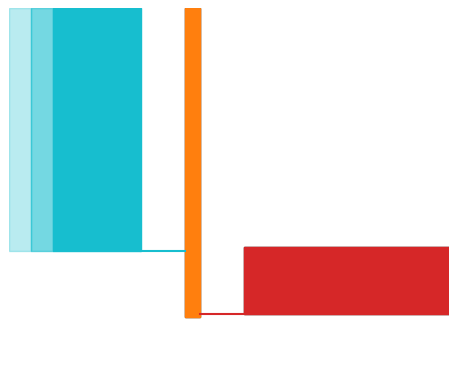} 
    \end{minipage}\hfill
    \begin{minipage}{0.45\textwidth}
        \centering
        \includegraphics[width=\textwidth]{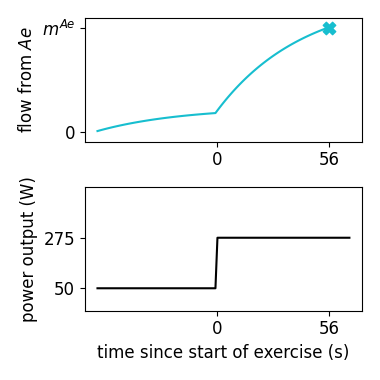}
    \end{minipage}}
    \caption{Left: A three component hydraulic model fitted to \gls{cp} 211~W and \gls{w'} 13240~J of participant~1. Values of all fitted parameters are in \Cref{tab:hyd_results} in the Appendix. Right: Flow from $Ae$ when the left model simulates a warm-up at 50 W and then a constant high intensity exercise at 275 W until maximal flow through $Ae$ is reached (x-symbol).}
    \label{fig:weig_sim}
\end{figure}

Due to our fitting procedure that adjusts hydraulic model parameters freely, it is not guaranteed that \gls{hyd_weig} conforms to the above discussed constraints on the \gls{m-m} model. As reported in our results in \Cref{sec:results}, the average parameter $\gamma$ of the fitted \gls{hyd_weig} was $0.017 \pm 0.005$ and the average $\phi$ was $0.263 \pm 0.038$. These values indicate that none of the \gls{hyd_weig} models adhered the constraint of \citet{morton_modelling_1990} that $\gamma > \phi$. This had a direct impact on \gls{vo2} predictions of \gls{hyd_weig}. As an example, on the left in \Cref{fig:weig_sim} is the system of tanks fitted to participant 1. Here, the parameters are $\gamma = 0.02$ and $\phi = 0.27$, so $\gamma < \phi$. In addition, the top of the right tank is much lower than the tank of the model depicted in \Cref{fig:m_m_sim}. The second constraint of \citet{morton_modelling_1990} suggests that the top of $AnS$ to be at approximately 40\% of the height of $Ae$, which is also not satisfied with the fitted model in \Cref{fig:weig_sim}. As a result, \gls{vo2} predictions of the model fitted to participant 1 did not resemble the \gls{vo2} slow component. The onset of flow from the right tank had almost no impact on the exponential increase of flow from the left tank. This explains the prediction errors in \Cref{tab:vo2_detail} and confirms on the example of \gls{vo2} predictions that the \gls{hyd_weig} model allows for unrealistic predictions for metabolic responses during exercise. 

\subsection{Warm-up \gls{vo2} predictions}

Another example of unrealistic \gls{vo2} predictions occurs at the beginning of exercise during the warm-up. Because the middle tank of \gls{hyd_weig} is filled initially, it will always predict the complete absence of \gls{vo2} at the first time step of a simulation. As observable in \Cref{fig:tte,fig:m_m_sim,fig:weig_sim}, this caused unrealistic \gls{vo2} predictions at the beginning of exercise tests. Predicted \gls{vo2} started at 0 and increased slowly throughout the 3~min of warm-up at 50~W. The size of the middle tank is fixed for the \gls{m-m} and \gls{hyd_weig} models. Therefore, if the flow from $Ae$ is interpreted as \gls{vo2}, these models are guaranteed to predict the absence of \gls{vo2} at the beginning of exercise and when the athlete is fully recovered.

\citet{lidar_validity_2021} acknowledged this issue and introduced a new parameter to shrink the size of the middle tank. Using the notation in \Cref{fig:weigend_old}, \citet{lidar_validity_2021} made the top of $AnF$ adjustable and introduced a parameter $\psi$, which defines the distance between the top of $AnF$ and the top of $Ae$. The closer $\psi$ to $1-\phi$, the larger the flow from $Ae$ at the start of the simulation. This additional parameter is not present in the \gls{m-m} or \gls{hyd_weig} models and is another argument for why \gls{hyd_weig} cannot make realistic predictions for \gls{vo2}.

\subsection{New labels for \gls{hyd_weig}}

\begin{figure}
    \centering
    \includegraphics[width=\textwidth]{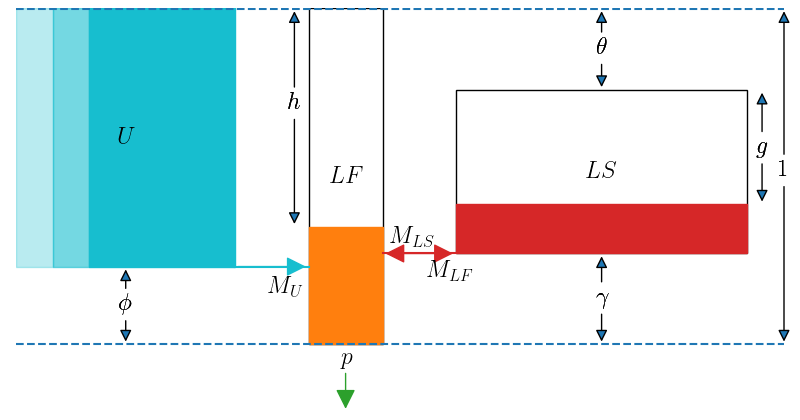} 
    \caption{We assigned new labels to our three component hydraulic model \gls{hyd_weig}. Tanks are named unlimited energy source $U$, limited fast energy source ($LF$), limited slow energy source ($LS$). Maximal flow capacities from these tanks are named $M_U$, $M_{LS}$, and $M_{LF}$.}
    \label{fig:new_notation}
\end{figure}

When the \gls{hyd_weig} model was first proposed in \cite{weigend_new_2021}, we chose to name the left infinitely big tank the aerobic energy source ($Ae$) and the two limited tanks the anaerobic fast ($AnF$) and anaerobic slow energy source ($AnS$). These names have a bioenergetic connotation and imply that the flow from the aerobic source still represents \gls{vo2} uptake like it did in the \gls{m-m} model and the model of \cite{margaria_biomechanics_1976}. The results of this work show that \gls{hyd_weig} does not predict realistic \gls{vo2} dynamics and should not be used for \gls{vo2} predictions.

Therefore, it is a sensible clarification to change the labels of \gls{hyd_weig} to more abstract names. The model with all new labels is depicted in \Cref{fig:new_notation}. Instead of $Ae$, we re-labelled the left tank as an unlimited energy source ($U$). Instead of $AnF$, the middle tank is now a limited fast energy source ($LF$), and instead of $AnS$, the right tank is a limited slow energy source ($LS$). These changes do not apply to any other hydraulic models and only affect \gls{hyd_weig}. They are an important step to protect users from misinterpreting fitted \gls{hyd_weig} models and their prediction results.

\section{Future work}

The above discussion recommends sensible directions for future work to broaden the range of possible applications for fitted three tank hydraulic models. Unrealistic \gls{vo2} predictions during high-intensity constant exercise can be improved by forcing a fitting approach to adhere to the previously discussed constraints of the \gls{m-m} model. \citet{lidar_validity_2021} did not consider these constraints and their fitted hydraulic models violated the same constraint that \gls{hyd_weig} violated in this work (in their Table 3 the $\lambda$ parameter of their model is smaller than $\phi$). Incorporating these constraints will considerably complicate the fitting procedures. Nevertheless, this could be a worthwhile direction for future research investigating if predictions of fitted three tank hydraulic models for metabolic responses during exercise can be improved.

However, the current focus and strength of \gls{hyd_weig} are energy recovery predictions during intermittent exercise, where promising new investigation opportunities emerge. Platforms like Strava\footnote{\url{https://www.strava.com}} or Golden Cheetah\footnote{\url{https://www.goldencheetah.org}} provide constantly growing databases of real-world intermittent exercise training and competition data. With affordable power metres, smartwatches, and online cycling apps such as Zwift\footnote{\url{https://www.zwift.com}}, the interest in intermittent exercise performance models and the amount of available data will grow. The future holds exciting possibilities for new iterations of \gls{w'bal} or hydraulic models, and the earlier we begin to develop pathways to investigate these models on such real-world data, the better. Therefore, while future work to apply hydraulic models to metabolic predictions and other modes of exercise is important, it is equally important to foster and develop its real-world application in intermittent exercise.

\section{Conclusion}

As already summarised by \cite{morton_critical_2006} for bioenergetic models or by \cite{skiba_w_2021} for \gls{w'bal} models, one-size-fits-all performance models do not exist. For simplicity, let us imagine a spectrum where simple and applicable performance models are on the left and complex and theoretical models are on the right. One could say that the critical power model is on the far left because it has only two parameters and requires just a few performance tests to be applied. Then, the hydraulic models of \cite{morton_critical_2006,lidar_validity_2021} and \cite{sundstrom_bioenergetic_2016} would be on the right because they have eight or more parameters and could not be applied to athletes due to the required in-depth knowledge about bioenergetic capacities. We showed that our \gls{hyd_weig} bridges the gap and remains somewhere in the middle. It can be applied to athletes and outperforms \gls{w'bal} models in intermittent exercise predictions. However, it should not be used for predictions of \gls{vo2} or alactic or lactic energy sources as the original hydraulic models were intended to be used. 

Depending on focus and setup, users must make well-informed decisions which of the available performance models are suitable for their analysis. This work further contributed to positioning our \gls{hyd_weig} model with its advantages and limitations among other existing performance models. 
We strive to promote progress in performance model development and to help users to make informed decisions for their analysis. As outlined in future work, there is an abundance of directions for creating new models and improving existing ones. To support this pursuit, we introduced new labels for our model. Furthermore, we embedded all our data, code of compared models, and hydraulic model advances in the open-source python packages \texttt{pypermod}\footnote{\url{https://github.com/faweigend/pypermod}} and \texttt{threecomphyd}\footnote{\url{https://github.com/faweigend/three_comp_hyd}}.

\section{Statements and Declarations}

\paragraph{Funding}
No funding was received to assist with the preparation of this manuscript.

\paragraph{Competing interests}
The authors have no competing interests to declare that are relevant to the content of this article.

\paragraph{Author contribution statement}
% Authors must provide a short description of the contributions made by each listed author (please use initials). This will be published in a separate section in front of the Acknowledgments.
FCW and JS and OO and EG conceived and designed research. FCW conducted data collection. FCW analysed the data. FCW and EG wrote the manuscript. All authors read and approved the manuscript.

\paragraph{Data and code availability}
Most relevant measurements on participants are summarised in the Appendix. Further, the datasets generated during the current study and the code for its analysis are available in the \texttt{pypermod} repository, \url{https://github.com/faweigend/pypermod}

%\begin{acknowledgements}
%\end{acknowledgements}

\newpage

\bibliographystyle{apalike}
\bibliography{references}

\newpage
\appendix
\section{Appendix}

\setcounter{table}{0}
\renewcommand{\thetable}{A\arabic{table}}
\setcounter{figure}{0}
\renewcommand{\thefigure}{A\arabic{figure}}

\begin{table}[h]
\caption{
An overview of critical power model fitting results for all participants. SEE\% denotes the standard error associated with the parameter as a percentage of the parameter, e.g, SEE\% of \gls{cp} is the \gls{SEE} associated with \gls{cp} divided by \gls{cp}.
} 
\begin{adjustwidth}{-0.75in}{-0.75in}\centering 
\begin{tabular}{ c c c c c c} 
\toprule
 & & \multicolumn{2}{c}{\gls{cp}} & \multicolumn{2}{c}{\gls{w'}} \\  
 \cmidrule(l){3-4}\cmidrule(l){5-6}
participant & best fit model & W & SEE\% & J & SEE\% \\
\midrule
1 & linear power-1/time & 211 & 3.1 & 13240 & 8.7 \\
2 & linear power-1/time & 292 & 2.4 & 19143 & 8.0 \\
3 & linear power-1/time & 238 & 0.7 & 10820 & 3.6 \\
4 & linear power-1/time & 199 & 1.8 & 16790 & 6.3 \\
5 & linear power-1/time & 174 & 1.5 & 14469 & 3.8 \\
\midrule
avg $\pm$ std & & $223 \pm 40$& $1.9 \pm 0.8$ & $14892 \pm 2869$ & $6.1 \pm 2.1$\\
\bottomrule
\end{tabular}
\end{adjustwidth}
\label{tab:cp_fitting_results}
\end{table}

\begin{table}[h]
\caption{An overview of parameters of fitted \gls{hyd_weig} models.}
\begin{adjustwidth}{-0.75in}{-0.75in}\centering 
\begin{tabular}{ c c c c S[table-format=3] S[table-format=3] S[table-format=1.2] S[table-format=1.2] S[table-format=1.2] } 
\toprule
participant& 
$AnF$ (J)& 
$AnS$ (J)& 
$m^{Ae}$ (W)& 
\multicolumn{1}{c}{$m^{AnS}$ (W)}& 
\multicolumn{1}{c}{$m^{AnF}$} (W)& 
$\theta$ &
$\gamma$ &
$\phi$ \\
\midrule
1 & 12562 & 36679 & 210 & 82 & 14 & 0.71 & 0.02 & 0.27 \\
2 & 18245 & 50537 & 291 & 117 & 20 & 0.7 & 0.02 & 0.27 \\
3 & 11914 & 38269 & 238 & 71 & 11 & 0.79 & 0.02 & 0.2 \\
4 & 16196 & 37141 & 198 & 95 & 15 & 0.68 & 0.01 & 0.27 \\
5 & 12733 & 30250 & 173 & 87 & 14 & 0.64 & 0.02 & 0.31 \\
\midrule
avg $\pm$ std & $14330 \pm 2463$& $38575 \pm 6605$& $222 \pm 40$& 
\multicolumn{1}{c}{$90 \pm 15$}& 
\multicolumn{1}{c}{$15 \pm 3$} & 
\multicolumn{1}{c}{$0.7 \pm 0.05$} & 
\multicolumn{1}{c}{$0.02 \pm 0.01$} & 
\multicolumn{1}{c}{$0.26 \pm 0.04$}\\
\bottomrule
\end{tabular}
\end{adjustwidth}
\label{tab:hyd_results}
\end{table}

\begin{figure}[h]
    \centering
    \includegraphics[width=\textwidth]{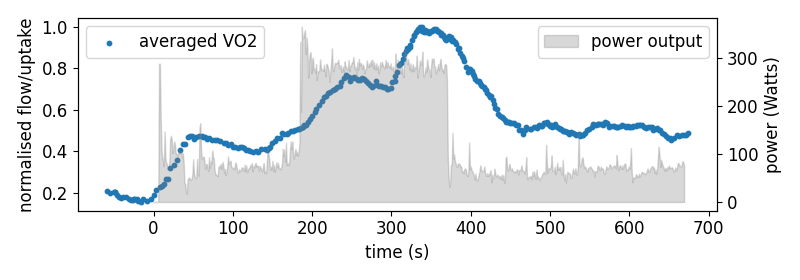}
    \caption{Measured power output (grey area) and \gls{vo2} uptake (dots) of participant~4 during a constant intensity exercise trial at 288 W. \gls{vo2} drops occurred because the face mask leaked air when the athlete lowered their head too far.}
    \label{fig:vo2-03}
\end{figure}

\begin{figure}[h]
    \centering
    \includegraphics[width=\textwidth]{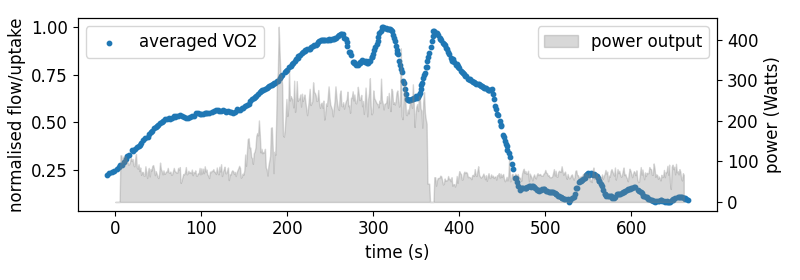}
    \caption{Measured power output (grey area) and \gls{vo2} uptake (dots) of participant~5 during a constant intensity exercise trial at 255 W. \gls{vo2} drops occurred because the face mask leaked air when the athlete lowered their head too far.}
    \label{fig:vo2-04}
\end{figure}
\end{document}